\documentclass[12pt]{article}
\input{psfig.sty}
\newcommand{\gtrsim}{ \mathop{}_{\textstyle \sim}^{\textstyle >} }
\newcommand{\lesssim}{ \mathop{}_{\textstyle \sim}^{\textstyle <} }

\newcommand{\vev}[1]{{\langle #1 \rangle}}
\newcommand{\abs}[1]{{\left| #1 \right|}}

\newcommand{\sL}{\mbox{$\widetilde L$}}
\newcommand{\HI}{\mbox{$H_{\rm inf}$}}
\newcommand{\mnu}{\mbox{$m_{\nu_1}$}}
\newcommand{\eV}{\mbox{~eV}}

\newcommand{\GeV}{\mbox{~GeV}}
\newcommand{\TeV}{\mbox{~TeV}}
\begin{document}
\baselineskip 0.6cm
%
\begin{titlepage}
\begin{center}

\begin{flushright}
DESY 01-142\\
\end{flushright}

\vskip 2cm

{\large \bf Affleck-Dine Leptogenesis and Low Scale Inflation}

\vskip 1.2cm

{T. Asaka}

\vskip 0.4cm

{\it Deutsches Elektronen-Synchrotron DESY, Hamburg, Germany}

\vskip 0.2cm

(October 4, 2001)

\vskip 2cm

\vskip .5in
\begin{abstract}
We study Affleck-Dine leptogenesis via the $\sL H_u$ flat direction in
supersymmetric theories.  We find that the baryon asymmetry is enhanced
when the energy scale of the inflation is sufficiently low.  Especially,
we consider models of low scale inflation in which the Hubble parameter
during inflation is comparable to (but slightly larger than) the
gravitino mass $m_{3/2} \sim 1 \TeV$.  The observed cosmological baryon
asymmetry is obtained with the lightest neutrino mass $m_{\nu_1} \sim
10^{-4}$ eV, if the reheating process is suddenly terminated after
inflation.
\end{abstract}
\end{center}
\end{titlepage}
\renewcommand{\thefootnote}{\arabic{footnote}} \setcounter{footnote}{0}
%
%
The origin of cosmic baryon asymmetry is one of the most fundamental
problems in particle physics and cosmology.  Although various mechanisms
have been proposed to solve it so far, the mechanism proposed by Affleck
and Dine~\cite{Affleck:1985fy} is particularly attractive if
supersymmetry (SUSY) is the physics beyond the standard model.  In the
SUSY standard model there appear various flat directions in the vacuum
configuration which carry $B$ and/or $L$ charges in the SUSY limit.
Their non-trivial evolution in the early universe could generate
sufficiently large $B$ and/or $L$ densities to explain the observed
baryon asymmetry.

Especially, Affleck-Dine (AD) leptogenesis via the $\sL H_u$ flat
direction~\cite{Murayama:1994em} ($\widetilde{L}$ is a scalar component
of the lepton-doublet superfield) has attracted the attention
\cite{Dine:1996kz}--\cite{Fujii:2001sn}, since there
is now convincing evidence of neutrino oscillations and the suggested
tiny neutrino masses indicate the lepton-number violation in nature,
which is essential to leptogenesis~\cite{Fukugita:1986hr}.  
It is also noteworthy that in leptogenesis scenarios
the present baryon asymmetry is closely related to masses and mixings
of neutrinos.  

Furthermore, the $\sL H_u$ direction is very special among various flat
directions relevant for the AD mechanism, since we can avoid the serious
problem associated with $Q$-balls.  The spatial
instability in the coherent oscillation of the AD field might lead to
the formation of $Q$-balls and spoil the simple description of the AD
baryogenesis~\cite{Kusenko:1998si}.
In the $\sL H_u$ direction, however, the $Q$-balls are not formed 
since its potential is steeper than quadratic one (and hence there is no
instability) due to the absence
of radiative correction from gluino loops and also the large contribution
from the top Yukawa coupling~\cite{Enqvist:2000gq}.
\footnote{In this analysis we assume gravity-mediated models of SUSY breaking 
with a gravitino mass $m_{3/2} \sim 1 \TeV$.}

Recently, detailed analyses of the AD leptogenesis were performed in
Refs.~\cite{Asaka:2000nb,Fujii:2001zr} including the effects of
surrounding thermal plasma pointed out in
Refs.\cite{Dine:1996kz,Allahverdi:2000zd,Anisimov:2000wx}.  It was shown
in Ref.~\cite{Fujii:2001zr} that the resultant baryon asymmetry $n_B/s$
(the ratio of the baryon number density $n_B$ to the entropy density $s$
in the present universe) is determined mainly by the mass of the
lightest neutrino $\mnu$ and is almost independent on the reheating
temperature of inflation $T_R$ in a wide region of $T_R \simeq
10^5$--$10^{12}$ GeV, and hence the observed value $n_B/s \simeq
(0.4$--$1) \times 10^{-10}$~\cite{Groom:2000in} predicts $\mnu \simeq
(0.1$--$3) \times 10^{-9}$ eV.  They also pointed out that such a
ultralight neutrino can be tested in the future experiments of
neutrinoless double beta decay.

In these analyses, parameters of the inflation model are considered as
free parameters.  Although the dependence on $T_R$ was discussed, the
energy scale of inflation was just assumed to be sufficiently high so
that the inflation ends well before the AD field starts to oscillate,
which is a crucial time since the net lepton asymmetry produced by the
$\sL H_u$ direction is fixed at this time.  However, if one considers
inflation models which take place at relatively low energy scale, the
above assumption might break down.  In this letter, therefore, we
investigate the AD leptogenesis in the presence of such a low scale
inflation and show that the resultant baryon asymmetry is
enhanced.
\footnote{Although
the AD leptogenesis introducing $U(1)_{B-L}$ gauge symmetry was
discussed in Ref.~\cite{Fujii:2001sn}, we do not consider this
possibility here.}

Let us start by explaining the AD leptogenesis via the flat
directions $H_u = \widetilde{L}_i$~\cite{Murayama:1994em}.  Here we follow
the discussion in Refs.~\cite{Asaka:2000nb,Fujii:2001zr}.  In the
minimal SUSY standard model we can incorporate neutrino masses by
introducing the effective operators in the superpotential,
\begin{eqnarray}
 \label{SP}
 W = \frac{ 1 }{ 2 M_i } (L_i H_u) (L_i H_u)~.
\end{eqnarray}
Through the seesaw mechanism~\cite{SeeSaw} neutrinos obtain masses
\begin{eqnarray}
 m_{\nu_i} &=& \frac{ \vev{H_u}^2 }{ M_i }~
 = \sin^2 \beta \times ( 3 \times 10^{-7} ) \eV
  \left( \frac{ 10^{20} \GeV }{ M_i }\right)~,
\end{eqnarray}
where $\vev{H_u}= \sin \beta \times 174$ GeV and we take $\sin \beta
\simeq 1$ since the final result does not change much.
\footnote{$\tan \beta = \vev{H_u}/\vev{H_d}$, where $H_u$ ($H_d$)
are Higgs fields which couple to up (down) type quarks, respectively.}  
Notice that, although we do not specify here the origin of the operators
in Eq.~(\ref{SP}), the scales $M_i$ in the presence of heavy Majorana
neutrinos correspond to roughly their masses divided by
squared of the neutrino Yukawa couplings and hence can be larger than
the reduced Planck scale $M_\ast = 2.4 \times 10^{18}$ GeV.  Since the
leptogenesis works most effectively for the flat direction of the first
family, we suppress the family index $i$ and consider only the flat
direction $\phi/\sqrt{2} \equiv H_u = \widetilde{L}_1$.  The flat
direction $\phi$ obtains its potential from SUSY breaking effects and
also from the non-renormalizable operator in Eq.~(\ref{SP}) as
\begin{eqnarray}
 \label{pot}
 V_0 = m_\phi^2 \abs{ \phi }^2
   + \frac{m_{3/2}}{8M} \left( a_m \phi^4 + h.c.\right)
   + \frac{ 1 }{ 4 M^2 } \abs{ \phi }^6,
\end{eqnarray}
where $m_\phi$ denotes the soft SUSY breaking mass
and $a_m$ is a coupling of order one.
We take $m_\phi \simeq m_{3/2} \simeq 1$ TeV and $\abs{a_m}\simeq 1$.

In the early universe, the potential (\ref{pot}) is modified as follows.
During the inflation and also the inflaton-oscillation period after the
inflation ends, the energy of the universe is dominated by the inflaton.
This non-zero energy induces an additional SUSY breaking which gives
corrections to the potential (\ref{pot})~\cite{Dine:1996kz}.  Although
the explicit form of these terms highly depends on details of the
K\"ahler potential and the inflation
model~\cite{Dine:1996kz,Gaillard:1995az}, we introduce here the
additional terms
\begin{eqnarray}
 \label{adpot}
 \delta V_{\rm inf} = - c_H H^2 \abs{ \phi }^2 + 
  \frac{ H }{ 8 M } \left( a_H \phi^4 + h.c. \right)~,
\end{eqnarray}
where $H$ denotes the Hubble parameter and
$c_H$ and $a_H$ are real and complex constants.
This is because certain values of $c_H$ and $a_H$ can explain 
the desirable initial condition for the AD mechanism.

Furthermore, $\phi$ receives an additional potential from the thermal
effects of the surrounding
plasma~\cite{Dine:1996kz,Allahverdi:2000zd,Anisimov:2000wx}.  It should
be noted that even in the period of the inflaton oscillation there is a
dilute plasma as a result of scatterings with the thermalized decay
products of the inflaton, which temperature is given by $T \simeq (
T_R^2 H M_\ast)^{1/4}$~\cite{Kolb-Turner}.  Here we do not explain these
thermal effects in detail but only give the induced terms
\begin{eqnarray}
 \label{thm}
 \delta V_{\rm th} = \sum_{f_k \abs{\phi} < T}  c_k f_k^2 T^2 \abs{\phi}^2
   + a_g \alpha_s^2 T^4 \log \left( \frac{ \abs{\phi}^2 }{ T^2 } \right)~,
\end{eqnarray}
where $f_k$ correspond to Yukawa or gauge coupling constants of the
field $\psi_k$ which couple to $\phi$ and $c_k$ are real positive
constants of order one. $\alpha_s$ is a strong coupling constant and
$a_g$ is a constant which is a bit larger than unity.  (Details can be
found in Refs.~\cite{Asaka:2000nb,Fujii:2001zr}.)

The effective total potential $V_{\rm tot}$, which is relevant for the
following discussion, is given by
\begin{eqnarray}
 \label{Vtot}
 V_{\rm tot} &=&
  \left( m_\phi^2 - c_H H^2 + \sum_{f_k \abs{\phi}< T} c_k f_k^2 T^2 
 \right) \abs{\phi}^2
  ~+~
  a_g \alpha_s^2 T^4 \log \left( \frac{ \abs{\phi}^2 }{ T^2 } \right)
\nonumber \\
 && ~+~  \frac{ m_{3/2} }{ 8 M } \left( a_m \phi^4 + h.c. \right)
 ~+~  \frac{ H }{ 8 M } \left( a_H \phi^4 + h.c. \right)
 ~+~ \frac{ \abs{\phi}^6 }{ 4 M^2 } ~.
\end{eqnarray}
With this potential we can describe the evolution of $\phi$ by the
equation of motion
\begin{eqnarray}
 \ddot{\phi} + 3 H \dot{\phi} 
  + \frac{ \partial V_{\rm tot} }{ \partial \phi^\ast} = 0~,
\end{eqnarray}
where the dot denotes a derivative with time.

During inflation the energy of the universe is dominated by
the vacuum energy of the inflaton and there is no thermal plasma.
The Hubble parameter takes an almost constant value $\HI$.
If $\HI$ is larger than $m_\phi$ and also 
$c_H \simeq \abs{a_H} \simeq 1$, 
it is found from Eq.~(\ref{Vtot}) that 
there is an instability of $\phi$ at origin and
$\phi$ is trapped at one of the minima of the potential
\begin{eqnarray}
 \abs{ \phi } &\simeq& \sqrt{ M \HI }, \\
 \arg{ \phi } &\simeq&
  \frac{ - \arg a_H + ( 2 n + 1) \pi }{ 4 } ~~(n = 0, 1, 2, 3)~.
\end{eqnarray}
This is because the curvature of the potential around the minimum along both
radius and phase directions is of the order of $\HI$, $\phi$ moves
towards one of the above minima from any given initial value and 
settles there.  This gives the desirable initial condition 
for the AD mechanism. Hereafter, 
we assume $c_H \simeq \abs{a_H} \simeq 1$ and consider only the
inflation models with $\HI \gtrsim m_\phi \simeq m_{3/2} \simeq 1 \TeV$.

After inflation ends, the energy of the universe is dominated by the
coherent oscillation of the inflaton until the reheating process
completes.  In this period, although there exists a dilute plasma, as
long as the potential for $\phi$ is dominated by $\delta V_{\rm
inf}$~(\ref{adpot}) and also $\abs{\phi}^6$ term in Eq.~(\ref{pot}), the
flat direction $\phi$ tracks the instantaneous minimum of the potential
\begin{eqnarray}
 \abs{ \phi } &\simeq& \sqrt{ M H }, \\
 \arg{ \phi } &\simeq&
  \frac{ - \arg a_H + ( 2 n + 1) \pi }{ 4 }~.
\end{eqnarray}
Therefore, the amplitude $\abs{\phi}$ decreases 
as $\abs{\phi} \propto H^{1/2} \propto t^{-1/2}$.

As the universe evolves the negative mass term (i.e., $- H^2
\abs{\phi}^2$) is eventually exceeded by another term in the full
potential (\ref{Vtot}).  At this time the evolution of $\phi$ is
drastically changed and $\phi$ begins to oscillate and to rotate around the
origin $\phi$ =0.  The Hubble parameter at this time $H_{\rm osc}$,
which is crucial to estimate the lepton asymmetry produced by $\phi$
(see below), is given by~\cite{Asaka:2000nb,Fujii:2001zr}
\begin{eqnarray}
 \label{Hosc}
  H_{\rm osc}
  ~\simeq ~\mbox{max}
  \left[ ~
   m_\phi,~
   H_k,~
   \alpha_s T_R \left( \frac{a_g M_\ast}{M} \right)^{1/2}~
 \right]~,
\end{eqnarray}
where $H_k$ are
\begin{eqnarray}
 H_k ~\simeq ~
  \mbox{min}
  \left[ ~
   \frac{M_\ast T_R^2}{f_k^4 M^2},~
   \left( c_k^2 f_k^4 M_\ast T_R^2 \right)^{1/3} ~
  \right]~.
\end{eqnarray}
It should be noted that $H_{\rm osc}$ should be smaller than $\HI$.
We shall assume this nontrivial fact for a while 
(see, however, the later discussion).

The evolution of $\phi$ for $H < H_{\rm osc}$ is fixed depending on
which term is dominated the total potential (\ref{Vtot}).  There are
three possibilities, i.e., the dominant term is (i) $m_\phi^2
\abs{\phi}^2$ term, (ii) $T^2 \abs{\phi}^2$ term, or (iii) $T^4 \log
(\abs{\phi}^2)$ term.  In each case, the damping rate of the amplitude
is estimated as (i) $\abs{\phi} \propto t^{-1}$, (ii)
$\abs{\phi} \propto t^{-7/8}$~\cite{Asaka:2000nb}, or
(iii) $\abs{\phi} \propto t^{-\alpha}$ with $\alpha
\simeq 1.5$~\cite{Fujii:2001zr}, respectively.
Note that the damping rate in all the above cases is faster than
the rate before $\phi$ starts to oscillate.

Now, we are at the point to estimate the lepton asymmetry produced 
in the considering AD leptogenesis.  
The lepton number density is given by
\begin{eqnarray}
 n_L = \frac{i}{2} 
  \left( \dot{\phi}^\ast \phi - \phi^\ast \dot{\phi^\ast} \right)~,
\end{eqnarray}
and its evolution is described by the equation
\begin{eqnarray}
 \label{EVOnL}
 \dot{n_L} + 3 H n_L =
  \frac{ m_{3/2} }{2 M} \mbox{Im} \left( a_m \phi^4 \right)
  +
  \frac{ H }{2 M} \mbox{Im} \left( a_H \phi^4 \right)~.
\end{eqnarray}
Notice that inflation sets $n_L = 0$ initially.  The phase of $\phi$
is kicked by the relative phase between $a_m$ and $a_H$ and the
rotational motion of $\phi$ generates the lepton
asymmetry~\cite{Affleck:1985fy}.  It was shown in
Refs.~\cite{Asaka:2000nb,Fujii:2001zr} that, comparing two source terms
of RHS in Eq.~(\ref{EVOnL}), the first term (i.e., the original $A$-term
in $V_0$ (\ref{pot})) gives the dominant contribution in generating
$n_L$.  Thus, by integrating Eq.~(\ref{EVOnL}) we obtain the produced
lepton asymmetry at time $t > t_{\rm osc} \sim H_{\rm osc}^{-1}$ as
\begin{eqnarray}
 \left[ R^3 n_L \right](t)
  &\simeq& \int^{t}
  dt' R^3 \frac{ m_{3/2} }{ 2 M } \mbox{Im}  \left( a_m \phi^4 \right)
\nonumber   \\
  &=& 
  \int^{t_{\rm osc}}
  dt' R^3 \frac{ m_{3/2} }{ 2 M } \mbox{Im}  \left( a_m \phi^4 \right)
  ~+~
  \int^{t}_{t_{\rm osc}}
  dt' R^3 \frac{ m_{3/2} }{ 2 M } \mbox{Im}  \left( a_m \phi^4 \right)~,
  \label{InL}
\end{eqnarray}
where $R$ is the scale factor of the universe.
The second term of RHS in Eq.~(\ref{InL}) 
gives only a small contribution to the total lepton asymmetry,
since $\mbox{Im}\left( a_m \phi^4 \right)$ changes its sign rapidly
due to the $\phi$ oscillation and also the damping rate
of $\abs{\phi}^4$ is faster than the rate of $R^{-3}$.
On the other hand, the integrand of the first term in Eq.~(\ref{InL})
is almost constant ($\propto t^0$) since $\abs{\phi} \propto t^{-1/2}$.
Therefore, the resulting lepton asymmetry for $t > t_{\rm osc}$ 
is dominated by the contribution at $H \simeq H_{\rm osc}$ and we have
\begin{eqnarray}
 n_L (t) \simeq n_L (t_{\rm osc}) \times 
  \frac{ R(t_{\rm osc})^3 }{ R(t)^3}
  \simeq     
  \frac{1}{3}
  \delta_{\rm eff}
  \abs{a_m}  m_{3/2} M H_{\rm osc}
  \times 
  \frac{ R(t_{\rm osc})^3 }{ R(t)^3}~,
\end{eqnarray}
where $\delta_{\rm eff} \simeq \sin ( 4 \mbox{arg} \phi + \mbox{arg}
a_m)$ denotes an effective CP-violating phase.

\begin{figure}[t]
    \vspace{-1cm}
    \centerline{\psfig{figure=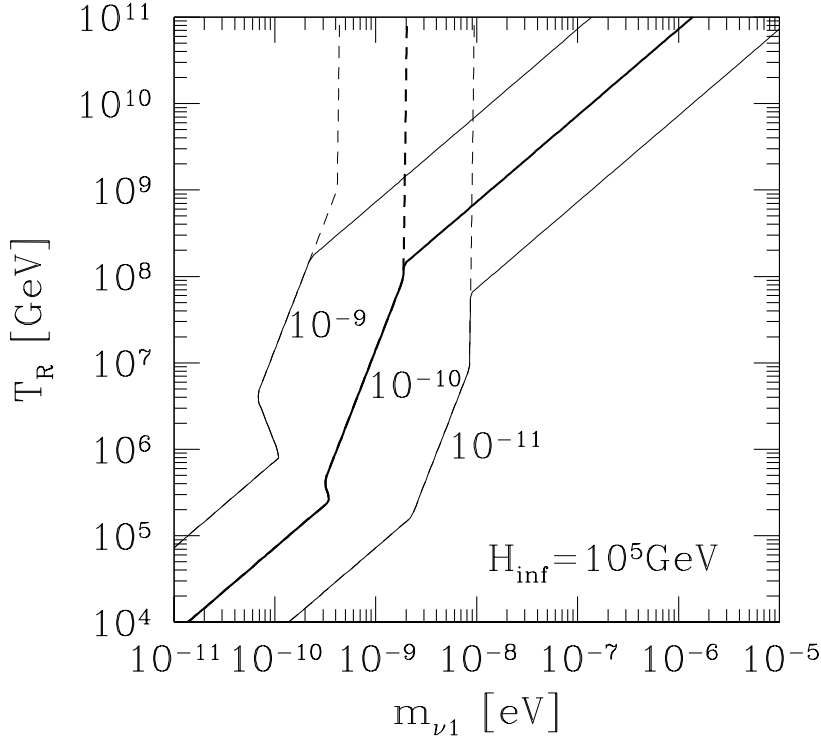,width=12cm}}
    \vspace{-1cm}
    \caption{
 Contour plot of the baryon asymmetry $n_B/s$ in the $m_{\nu_1}$--$T_R$
 plane. The lines represent the contour plots for $n_B/s = 10^{-9}$,
 $10^{-10}$, and $10^{-11}$ from the left to the right. 
 The dashed lines represent the ones for the inflation models with
 $\HI > H_{\rm osc}$. The solid lines represents the ones when
 $\HI=10^5$ GeV.  We take $\abs{\delta_{\rm eff}} = 1$.}
    \label{fig:BA}
\end{figure}

\begin{figure}[t]
    \vspace{-1cm}
    \centerline{\psfig{figure=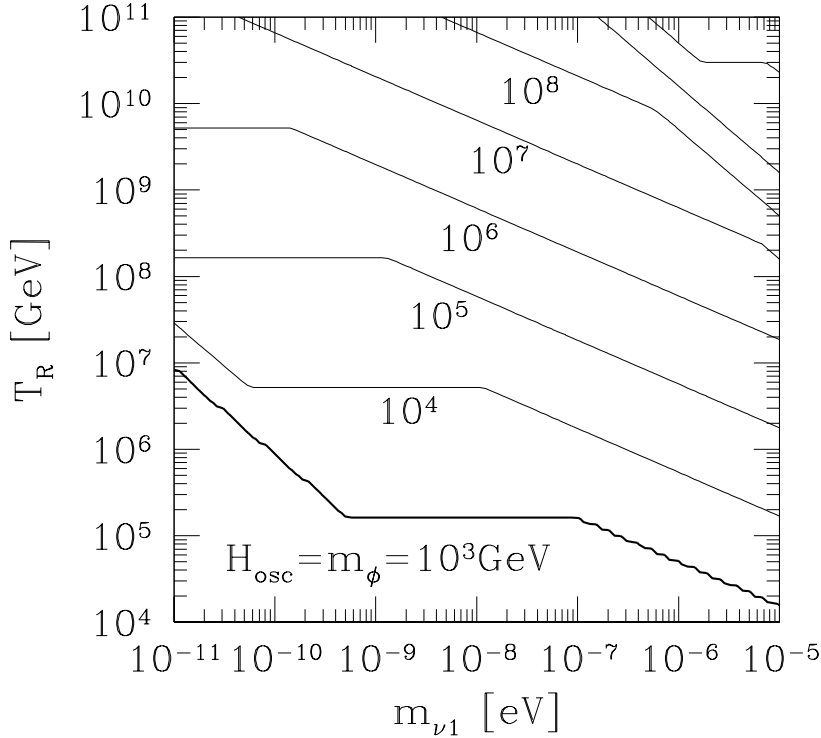,width=12cm}}
    \vspace{-1cm}
    \caption{
 Contour plot of the Hubble parameter $H_{\rm osc}$ in the
 $m_{\nu_1}$--$T_R$ plane.  Corresponding values of $H_{\rm osc}$ are
 represented in unit of GeV.  In the region below the thick line $H_{\rm
 osc}$ takes a constant value $H_{\rm osc} = m_\phi = 1 \TeV$.  }
    \label{fig:Hosc}
\end{figure}

The lepton-to-entropy
ratio when the reheating process of inflation completes at $T = T_R$ is
estimated as
\begin{eqnarray}
 \label{nL}
 \frac{n_L}{s} \simeq \delta_{\rm eff} ~
  \frac{ M T_R }{ 12 M_\ast^2 }
  \left( \frac{ m_{3/2} }{ H_{\rm osc}}\right)~.
\end{eqnarray}
This ratio takes a constant value as long as no dilution exists 
in the later epoch.
The lepton asymmetry, since it is produced well before 
the electroweak phase transition (at $T \sim 10^2 \GeV$),
is partially converted into
the baryon asymmetry
from the chemical equilibrium between lepton and baryon number
through the sphaleron effects~\cite{Kuzmin:1985mm}.
\footnote{In the minimal SUSY standard model the relation between lepton
and baryon asymmetries is given by
$n_B= - (8/23) n_L $~\cite{Khlebnikov:1988sr}.
} 
The baryon asymmetry in the present universe
is~\cite{Asaka:2000nb,Fujii:2001zr}
\begin{eqnarray}
 \label{BA0}
 \frac{n_B}{s} =
  \abs{\delta_{\rm eff}} \frac{ 2 M T_R }{ 69 M_\ast^2}
  \left( \frac{ m_{3/2} }{ H_{\rm osc} }  \right)~.
\end{eqnarray}
Here we neglected the sign of the produced baryon asymmetry.  One 
sees that the AD leptogenesis is most effective for the flat direction of
the first family corresponding to the largest scale $M$ (or the lightest
neutrino mass $m_{\nu_1}$).  If $H_{\rm osc} \gg m_\phi \simeq m_{3/2}$
due to the early oscillation by the thermal effects, the baryon
asymmetry is suppressed.  In Fig.~\ref{fig:BA} we show the contour plot
of $n_B/s$ by the dashed lines.  It is found that the present baryon
asymmetry is determined almost independently on the reheating
temperature for a wide range of $T_R \simeq 10^5$--$10^{12} \GeV$ and
the observed value $n_B/s \simeq (0.4-1)\times 10^{-10}$ suggests an
ultralight neutrino with a mass $\sim {10^{-9}}$~eV~\cite{Fujii:2001zr}.

In deriving Eq.~(\ref{BA0}) we made two assumptions on parameters of the
inflation model.  First, we assumed that the reheating temperature is
sufficiently low and its process completes after the lepton asymmetry is
fixed at $H \simeq H_{\rm osc}$.  In the parameter space shown in
Fig.~\ref{fig:BA} this assumption is justified.  Moreover, such a low
reheating temperature is preferred to avoid the cosmological gravitino
problem (see the discussion below).

The second assumption is that the scale $v$ of the inflation is
sufficiently high and the AD field starts to oscillate well after the
inflation ends, i.e., $ H_{\rm inf} = v^2/(\sqrt{3} M_\ast) > H_{\rm
osc}$.  We show in Fig.~\ref{fig:Hosc} the contour plot of $H_{\rm
osc}$.  It is seen that the scale of $H_{\rm osc}$ becomes higher in the
heavier $m_{\nu_1}$ region due to the large thermal effects.  Therefore,
for a model of low scale inflation with a fixed $H_{\rm inf}$ this
assumption breaks down in some region of parameter space
and the final expression of the present
baryon asymmetry in Eq.~(\ref{BA0}) should be modified.

Thus, we discuss the AD leptogenesis for the case
$H_{\rm osc} > H_{\rm inf}$.
As mentioned before, we consider only the models with
$H_{\rm inf} \gtrsim m_\phi \simeq m_{3/2}$ to explain the initial
condition for the AD mechanism.  Since the energy of the radiation after
the reheating process completes is smaller than the inflaton vacuum
energy $v^4$, one has
\begin{eqnarray}
 \label{TRR}
 T_R < 0.5 \sqrt{ H_{\rm inf} M_\ast} 
  = 2 \times 10^{11} \GeV 
  \left( \frac{ H_{\rm inf} }{ 10^5 \GeV} \right)^{1/2}~.
\end{eqnarray}
Considering the evolution of the dilute plasma in the period
of the inflaton oscillation, the maximum temperature $T_{\rm MAX}$ after
the inflation is achieved when $H = H_{\rm MAX} = 0.6 H_{\rm inf}$
and is given by~\cite{Kolb-Turner}
\begin{eqnarray}
 T_{\rm MAX} \simeq \left( T_R^2 H_{\rm inf} M_\ast \right)^{1/4}~.
\end{eqnarray}
The temperature for $H < H_{\rm MAX}$ is given by $T \simeq ( T_R^2 H
M_\ast )^{1/4}$ as long as $T > T_R$.  Therefore, the previous
discussion can be applied for $H \le H_{\rm MAX}$.
Neglecting the small
time difference between $H_{\rm inf}$ and $H_{\rm MAX}$, it is found
that the AD field starts to oscillate and hence the produced lepton
asymmetry is fixed just after the inflation ends at $H \simeq H_{\rm
inf}$, since $H_{\rm osc} > H_{\rm inf}$.  The present baryon asymmetry
in the considering case is obtained by replacing $H_{\rm osc}$ by
$H_{\rm inf}$ in Eq.~(\ref{BA0}) as
\begin{eqnarray}
 \frac{n_B}{s} \simeq
  \abs{\delta_{\rm eff}} \frac{ 2 M T_R }{ 69 M_\ast^2}
  \left( \frac{ m_{3/2} }{ H_{\rm inf} }  \right)~.
\end{eqnarray}
In Fig.~\ref{fig:BA} we also show the contour plot of 
$n_B/s$ by the solid lines for the case
$H_{\rm inf}=10^5$ GeV (i.e., $v = 6 \times 10^{11} \GeV$).  It is
seen that the present baryon asymmetry is enhanced by the rate $H_{\rm
osc}/H_{\rm inf}$ in the region $H_{\rm osc} > H_{\rm inf}$.  In this
case, with relatively high reheating temperatures, the lightest neutrino
mass $m_{\nu_1} \sim 10^{-6}$ eV ($\gg 10^{-9}$ eV) is sufficient to
explain the observed baryon asymmetry.  Notice that such low values of $H_{\rm
inf}$ are available in a class of SUSY inflation
models~\cite{Izawa:1997dv,German:2000yz}.
\footnote{Here we do not specify the value of the reheating
temperature predicted by the inflation models, but take $T_R$ as a free
parameter in the region (\ref{TRR}).}

Finally, we consider the extreme case that the Hubble parameter is
comparable to (but slightly larger than) the SUSY breaking mass of the AD
field $\phi$ ($H_{\rm inf} \sim m_\phi \simeq m_{3/2}$).  In this case
$\phi$ starts to oscillate just after the end of inflation at $H \simeq
H_{\rm inf} \sim m_\phi$ and hence the produced lepton asymmetry is
determined independently on details of the additional potential
(\ref{thm}) induced by the thermal plasma.  Then, the resultant baryon
asymmetry is enhanced since there is no suppression coming from the
early oscillation by the thermal effects and its expression is given by
dropping off the factor $m_{3/2}/H_{\rm osc}$ in Eq.~(\ref{BA0}), which
is the one obtained without including the thermal effects in the earlier
works.\footnote{See, for example, Ref.~\cite{Moroi:2000uc}.}

Furthermore, if the reheating process completes just after
the end of the inflation, we expect to have a larger baryon asymmetry since
the lepton asymmetry produced at $H \simeq H_{\rm inf} \sim m_\phi$ does
not receive the entropy dilution by the inflaton decay.  In this sudden
reheating case, the inflationary epoch is just followed by the radiation
dominated universe and the reheating temperature is given by
\begin{eqnarray}
 T_R = 0.5 \sqrt{ \HI M_\ast}= 2 \times 10^{10} \GeV 
  \left( \frac{ \HI }{ m_{3/2} } \right)^{1/2}
  \left( \frac{ m_{3/2} }{ 1 \TeV } \right)^{1/2}~.
\end{eqnarray}
At this time the lepton asymmetry produced by $\phi$ is fixed as
\begin{eqnarray}
 \frac{n_L}{s} 
  \simeq
 \frac{ \frac{1}{4} \delta_{\rm eff} \abs{a_m} m_{3/2} \HI M }
      { \frac{2 \pi^2}{45} g_\ast T_R^3}~,
\end{eqnarray}
where we used the fact $H = 1/(2t)$ in the radiation dominated universe.
We obtain, then, the present baryon asymmetry 
\begin{eqnarray}
 \frac{n_B}{s} \simeq 
  3 \times 10^{-11} \delta_{\rm eff} \abs{a_m} 
  \left( \frac{m_{3/2}}{\HI} \right)^{1/2}
  \left( \frac{m_{3/2}}{1\TeV} \right)^{1/2}
  \left( \frac{10^{-4}\eV}{\mnu}\right)~.
\end{eqnarray}
Therefore, for models of low scale inflation with 
$H_{\rm inf} \sim m_{3/2}$, if the reheating process
is suddenly terminated,  the lightest neutrino mass of 
$m_{\nu_1} \sim 10^{-4}$ eV is small enough to account for the observed
baryon asymmetry in the present universe.

Before closing this letter, we should mention the constraint on $T_R$
from the cosmological gravitino problem.  Recent
analysis~\cite{Kawasaki:2001qr}, using the gravitino density given in
Ref.~\cite{Bolz:2001fu},
\footnote{The upper bound on $T_R$ depends on a gluino mass 
$m_{\tilde g}$. We take $m_{\tilde g} =1$ TeV.} 
shows that the nucleosynthesis puts the
the upper bound on $T_R$ as $T_R \lesssim 10^{6}$, $10^{9}$ and
$10^{12}$ GeV for the gravitino mass $m_{3/2} = 100$ GeV, 1 TeV and 3
TeV, respectively.
\footnote{The gravitinos produced nonthermally in the preheating
epoch~\cite{NTHgra} are ignored.} 
Therefore, a gravitino mass of a few TeV is sufficient to have the
relatively high reheating temperatures in the present analysis.  Further,
considerably higher reheating temperatures are acceptable in some
cases~\cite{Asaka:2000ew}.

%
%
\section*{Acknowledgements}
The author would like to thank W.~Buchm\"uller for helpful discussions
and a careful reading of the manuscript, and K.~Hamaguchi for useful comments.
%
%

\end{document}